\documentstyle[eqsecnum,aps,prd,psfig,preprint]{revtex}
\begin{document}
\draft

\title{Gravitational field of vacuumless defects}
\author{Inyong Cho\footnote[1]{Electronic address: cho@cosmos2.phy.tufts.edu}
and Alexander Vilenkin\footnote[2]{Electronic address: vilenkin@cosmos2.phy.tufts.edu}}

\address{Institute of Cosmology,
        Department of Physics and Astronomy,\\
        Tufts University,
        Medford, Massachusetts 02155, USA}
\date{\today}
\maketitle

\begin{abstract}
It has been recently shown that topological defects can arise in
symmetry breaking models where the scalar field potential $V(\phi)$
has no minima and is a monotonically decreasing function of $|\phi|$.
Here we study the gravitational fields produced by such vacuumless
defects in the cases of both global and gauge symmetry breaking.
We find that a global monopole has a strongly repulsive gravitational
field, and its spacetime has an event horizon similar to that in 
de Sitter space.
A gauge monopole spacetime is essentially that of a magnetically
charged black hole. 
The gravitational field of a global string is repulsive and that of
a gauge string is attractive at small distances and repulsive at
large distances. Both gauge and global string spacetimes have singularities
at a finite distance from the string core.

\end{abstract}


\section{Introduction}
Topological defects are high-energy relics which could be formed
at symmetry-breaking phase transitions in the early 
universe~\cite{Kibble,Shell}.
A prototypical symmetry breaking model is described by the Lagrangian
\begin{equation}
{\cal L}={1 \over 2}\partial_\mu\phi_a \partial^\mu\phi_a - V(f)\;,
\label{lag}
\end{equation}
where $\phi_a$ is a set of scalar fields, $a=1,...,N$, 
$f=(\phi_a\phi_a)^{1/2}$, and the potential $V(f)$ has a minimum at
a non-zero value of $f$.
The model has O(N) symmetry and admits domain wall, string, and monopole
solutions for $N=1,2, \text{ and } 3$, respectively.

One can consider a different class of models in which $V(f)$ has a local
maximum at $f=0$ but no minima; instead, it monotonically decreases to zero
at $f\to\infty$. For example,
\begin{equation}
V(f)=\lambda M^{4+n}(M^n+f^n)^{-1}\;.
\label{pot}
\end{equation}
In the particle physics context, potentials without stable vacua can
arise due to non-perturbative effects in supersymmetric 
theories~\cite{Seiberg}.
Models of this sort have attracted considerable interest in cosmology,
where the ``rolling'' field $\phi$ may significantly contribute
to the present energy density of the universe~\cite{Peebles}.

In a recent paper~\cite{Cho} we have studied the properties of topological
defects in models of the type~(\ref{lag}),~(\ref{pot}), for both
gauge and global symmetries. Since the potential $V(f)$ does not have 
a stable vacuum and decreases monotonically with $f$, the scalar field
magnitude $f(r)$ grows unboundedly with the distance $r$ from the defect
core. We have introduced the term ``vacuumless'' for defects of this sort.
The defect solutions of Ref.~\cite{Cho} have been found in flat spacetime.
The purpose of present paper is to study the gravitational field
of vacuumless defects.

Let us first summarize the relevant conclusions of Ref.~\cite{Cho}.

\noindent
(i) For a vacuumless domain wall centered at $x=0$, the asymptotic
behavior of the scalar field and of the energy density is
$f(x)\propto |x|^{2/(n+2)}$ and $T_0^0(x)\propto |x|^{-2n/(n+2)}$
at $|x|\to\infty$.
For $n>2$, the energy distribution is integrable, with most of the energy
concentrated near the wall mid-section ($x=0$). 
We expect, therefore, that the gravitational field of vacuumless walls
will not be much different from that of ordinary domain walls.

\vspace{0.2in}
\noindent
(ii) For global vacuumless strings and monopoles, the flat-space solution is
\begin{equation}
f(r)=aM(r/\delta)^{2 \over n+2}\;,
\label{fgbsm}
\end{equation}
where
$\delta =\lambda^{-1/2}M^{-1}$ is the size of the defect core, $r$ is the
distance from the monopole center for monopoles and the distance from
the string axis for strings, and $a\sim 1$ is a numerical coefficient,
$a=(n+2)^{2/(n+2)}(n+4)^{-1/(n+2)}$ for strings and
$a=(n+2)^{2/(n+2)}[2(n+3)]^{-1/(n+2)}$ for monopoles.
The solution~(\ref{fgbsm}) applies for $\delta \ll r \ll R$,
where $R$ is the cutoff radius determined by the distance to the
nearest string or monopole.
Vacuumless strings and monopoles are very diffuse objects, with
most of their energy distributed at large distances from the core
($r\sim R$), and with potential and gradient energy contributing
comparable amounts. They are much more diffuse than ordinary strings
and monopoles, and we expect their spacetimes to be substantially
different from the ordinary case.

\vspace{0.2in}
\noindent
(iii) Gauge vacuumless strings have magnetic flux localized within
a thin tube inside the core, while outside the core the scalar field is
given by
\begin{equation}
f(r)=a\ln(r/\delta)+b\;.
\label{fggs}
\end{equation}
An unusual feature of this solution is that the coefficients $a$ and $b$
are sensitive to the cutoff distance $R$,
\begin{equation}
a\sim M(R/\delta)^{2 \over n+2}[\ln(R/\delta)]^{-{n+1 \over n+2}}\;,\quad
b\sim a\ln(R/\delta)\;.
\label{abgs}
\end{equation}
The energy distribution at small distances is dominated by the gradient
energy with $T_0^0\propto r^{-2}$, while the potential energy begins to
dominate at $r \gtrsim R/\ln^{1/2}(R/\delta)$.

\vspace{0.2in}
\noindent
(iv) Finally, a flat-space vacuumless gauge monopole is described by
the Prasad-Sommerfield solution. 
For $r\ll R$, the scalar field is given by
\begin{equation}
f(r)=a\coth (ear) -1/er\;,
\label{fggm}
\end{equation}
where
\begin{equation}
a\sim M(\lambda e M^3R^3)^{1/(n+1)}
\label{aggm}
\end{equation}
and $e$ is the gauge coupling. The size of the monopole core is
$r_0\sim 1/ea$. Much of the monopole energy is concentrated 
near the core; the total energy in that region is
\begin{equation}
{\cal M} = {4\pi \over e}a\;.
\label{mggm}
\end{equation}
In addition, there is a nearly constant energy density $T_0^0\approx V(a)$
outside the core. Its contribution to the total energy is comparable 
to~(\ref{mggm}), $V(r)R^3\sim {\cal M}$.

In the next section, the spacetimes of vacuumless defects will be
studied using the linearized gravity approximation.
Numerical solutions of the full Einstein's, scalar, and gauge field
equations will be discussed in Sec.~III.

\section{Linearized gravity}
We shall first consider the Newtonian approximation which applies for
weak gravitational fields and non-relativistic motion.
The Newtonian potential $\Phi$ can be found for the equation 
\begin{equation}
\nabla^2\Phi = 4\pi G(T_0^0 - T_i^i)\;.
\label{newteq}
\end{equation}
For a global defect described by the Lagrangian~(\ref{lag}),
\begin{equation}
T_\mu^\nu = \partial_\mu\phi_a\partial^\nu\phi_a - 
\delta_\mu^\nu {\cal L}\;,
\label{emtensor}
\end{equation}
and, for a static field $\phi_a({\bf {\rm x}})$,
\begin{equation}
T_0^0-T_i^i =-2V(f)\;.
\label{em2}
\end{equation}
For global vacuumless strings and monopoles, $f(r)$ is given by 
Eq.~(\ref{fgbsm}) and
\begin{equation}
T_0^0-T_i^i = -{2\lambda M^4 \over a^n} \left( {\delta \over r}
\right)^{2n \over n+2}\;.
\label{em3}
\end{equation}
The solution of Eq.~(\ref{newteq}) is then
\begin{equation}
\Phi (r) =-KGM^2(r/\delta)^{4/(n+2)}\;,
\label{Phi}
\end{equation}
where $K\sim 1$ is a numerical coefficient,
$K=2\pi (n+2)^2/a^n(n+6)$ for monopoles and
$K=\pi (n+2)^2/2a^n$ for strings.
One could also include a solution of the homogeneous equation,
$\Phi_0 (r)$, representing the gravitational field produced by
the defect core ($\Phi_0 \propto r^{-1}$ for monopoloes and
$\Phi_0 \propto \ln (r/\delta)$ for strings), but it becomes 
subdominant at sufficiently large $r$.

The potential~(\ref{Phi}) describes a strong repulsive gravitational
field. This is not surprising, considering the sign of the source in
Eq.~(\ref{em2}). Note that for $n=2$ the repulsive force is independent
of the distance from the defect core.
The linearized approximation applies as long as $\Phi (r) \ll 1$.
From~(\ref{fgbsm}) and~(\ref{Phi}) we see that this is equivalent to
$f(r) \ll m_p$, where $m_p = G^{-1/2}$ is the Planck mass.

For a gauge string, the gauge field is negligible outside the core,
the energy momentum tensor~(\ref{emtensor}) is modified by replacing
ordinary derivatives of $\phi_a$ by gauge-covariant derivatives,
and Eq.~(\ref{em2}) is still valid.
However, in this case the potential energy is negligible compared to
the gradient energy, provided that $r$ is not too large,
$r \ll R/\ln^{1/2}(R/\delta)$.
The gravitational potential in this range of $r$ should then be given
by a solution of the homogeneous equation, $\Phi(r)\propto \ln(r/\delta)$.

We now turn to linearized Einstein's gravity. The gravitational field
of a vacuumless global monopole is described by a static, 
spherically-symmetric metric which can be represented as
\begin{equation}
ds^2 = -B(r)dt^2 +A(r)dr^2 +r^2d\Omega^2\;,
\label{mmetric}
\end{equation}
where $d\Omega^2$ is the metric on a unit 2-sphere. Einstein's equations
for this metric are given in the Appendix. Expressing $A$ and $B$ as
\begin{equation}
A(r)=1+\alpha (r)\;, \quad B(r) =1 +\beta (r)\;,
\label{ABmon}
\end{equation}
linearizing in $\alpha$ and $\beta$,
and using the flat space expression~(\ref{fgbsm}) for $f(r)$,
the equations are reduced to
\begin{equation}
\beta '' +{2 \over r}\beta ' = -{16\pi\lambda \over a^n}
GM^4\left( {\delta \over r} \right)^{2n \over n+2}\;,
\label{betaeqmon}
\end{equation}
\begin{equation}
{\alpha ' \over r}+{\beta ' \over r} = 
-8\pi\lambda \left( {2 \over n+2} \right)^2 
a^2GM^4\left({\delta \over r}
\right)^{2n \over n+2}\;.
\label{alphaeqmon}
\end{equation}
The solution is easily found and can be expressed in the form
\begin{equation}
ds^2 = -(1+2\Phi)dt^2+(1+k\Phi)dr^2+r^2d\Omega^2\;,
\label{lrmmetric}
\end{equation}
where $\Phi(r)$ is the Newtonian potential~(\ref{Phi}) and
$k\sim 1$ is a numerical coefficient.
The coefficients $k$ and $K$ are functions of $n$ given by rather
complicated expressions which we shall not reproduce here.
In the special case of $n=2$ which we shall use in our numerical examples,
$a=2/10^{1/4}=1.13$, $K=3\pi$, and $k=17/16$.

For a global vacuumless string, we assume that the spacetime is static,
cylindrically-symmetric, and also has a symmetry with respect to
Lorentz boosts along the string axis.
With a suitable choice of coordinates, the metric can be written as
\begin{equation}
ds^2=B(r)(-dt^2+dz^2)+dr^2+C(r)d\theta^2\;.
\label{smetric}
\end{equation}
The corresponding Eisntein's equations are given in the Appendix.
In the weak gravity approximation,
$B(r)=1+\beta (r)$, $C(r)=r^2[1+\gamma (r)]$, and the linearized equations
for $\beta$ and $\gamma$ can be solved in the same way as for monopoles.
The resulting metric is
\begin{equation}
ds^2=(1+2\Phi)(-dt^2+dz^2)+dr^2+(1+k\Phi)r^2d\theta^2\;.
\label{lrsmetric}
\end{equation}
Once again, $\Phi (r)$ is the Newtonian potential~(\ref{Phi}) and
$k\sim 1$ is a numerical coefficient.
For $n=2$, we have $a=2/6^{1/4}=1.28$, $K=15.1$, 
and $k=1.7$.

For a gauge vacuumless string, the energy-momentum tensor for the
solution~(\ref{fggs}) is dominated by gradient terms,
\begin{equation}
T_0^0=T_z^z=T_\theta^\theta=-T_r^r={f'^2 \over 2}={a^2 \over 2r^2}\;.
\label{emgs}
\end{equation}
The metric can still be expressed in the form~(\ref{smetric}),
and the linearized Einstein's equations take the form
\begin{equation}
\beta '' + {1 \over r}\beta ' =0\;,
\label{gseq1}
\end{equation}
\begin{equation}
\beta '' -{1 \over r}\beta ' =8\pi G {a^2 \over r^2}\;,
\label{gseq2}
\end{equation}
\begin{equation}
\gamma '' + {2 \over r}\gamma ' + 2\beta ''=-16\pi G {a^2 \over r^2}\;.
\label{gseq3}
\end{equation}
The solution is easily found, and the resulting metric has the form
\begin{equation}
ds^2=(1+2\Phi)(-dt^2+dz^2)+dr^2+(1-4\Phi)r^2d\theta^2\;.
\label{lrgsmetric}
\end{equation}
Here,
\begin{equation}
\Phi = 2\pi Ga^2\ln(r/\delta)
\label{Phigs}
\end{equation}
and $a$ is given by Eq.~(\ref{abgs}).

\section{Numerical solutions}
To go beyond the linear approximation, we found the gravitational field
of vacuumless defects by numerically solving the combined Einstein's
and matter field equations.

\begin{figure}
\psfig{file=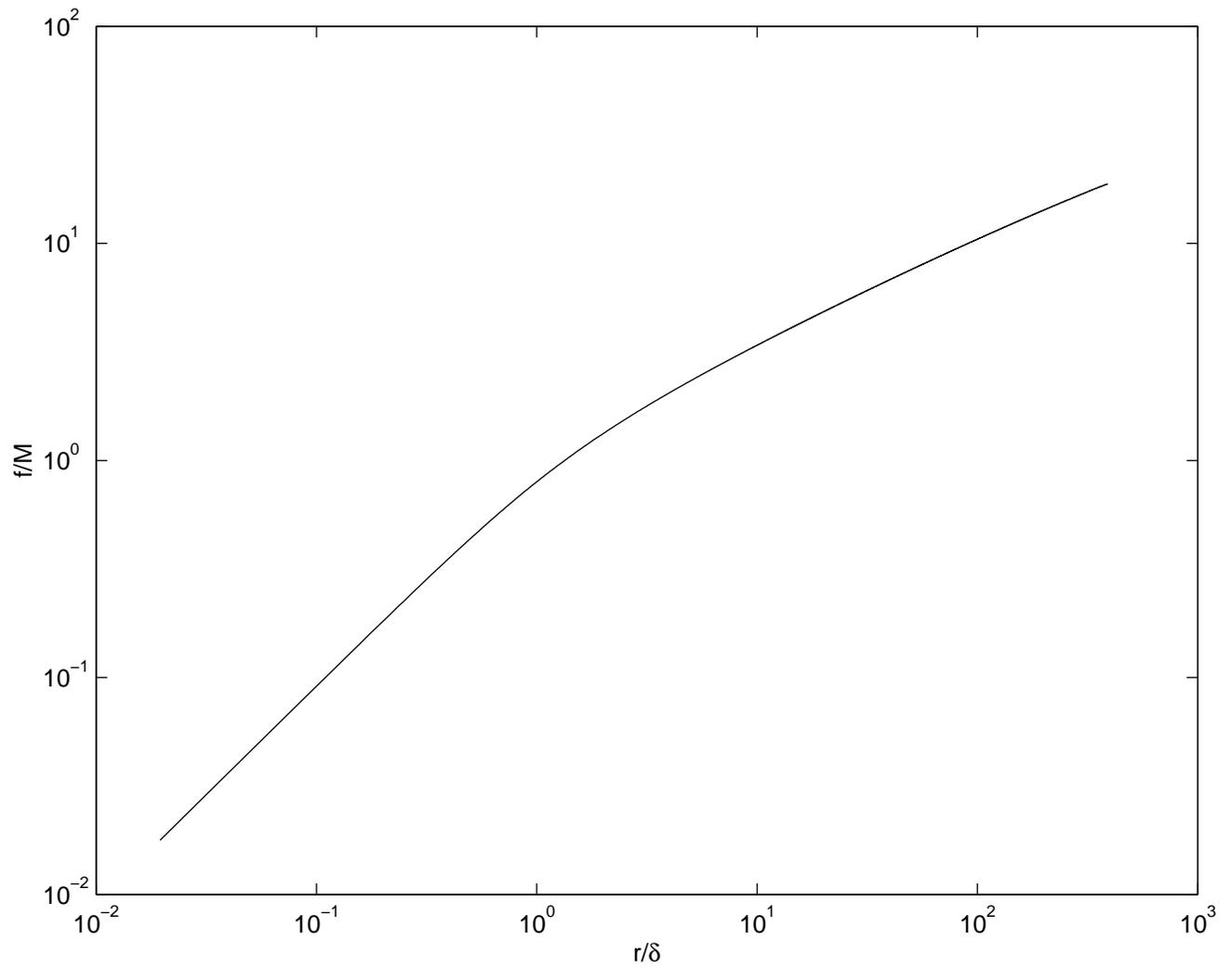}
\caption{
Scalar field $f(r)$ of a vacuumless global monopole
with $M=0.01m_p$.
}
\label{fig=gbmf}
\end{figure}

\begin{figure}
\psfig{file=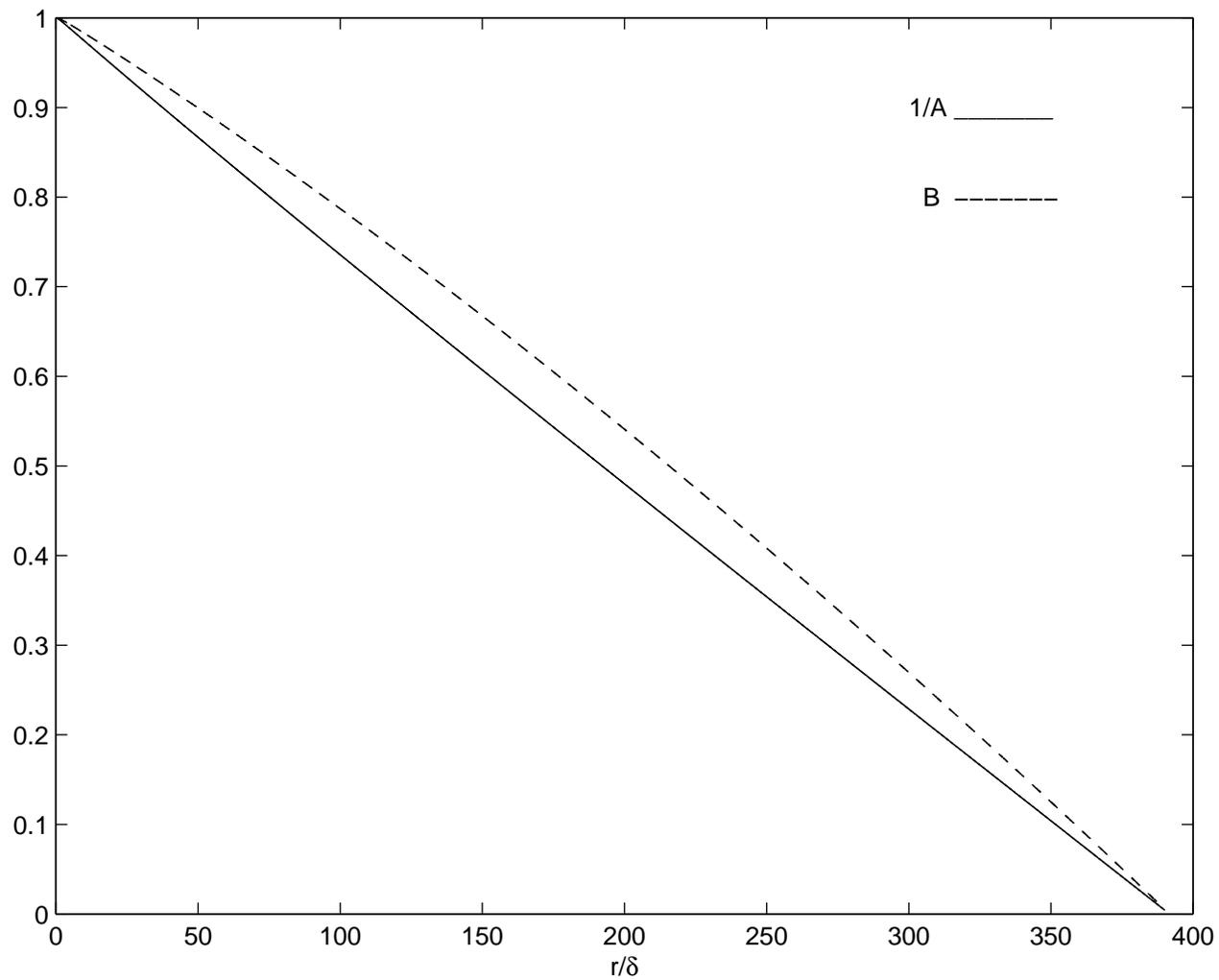}
\caption{
Metric coefficients $B(r)$ and $1/A(r)$ of a vacuumless global monopole
with $M=0.01m_p$.
}
\label{fig=gbm}
\end{figure}

For a global monopole, we used the metric ansatz~(\ref{mmetric})
with the boundary conditions
\begin{equation}
A(0)=B(0)=1\;,
\label{BCm}
\end{equation}
\begin{equation}
f(0)=0\;,\quad f'(R)=0\;.
\label{BCf}
\end{equation}
The parameters of the scalar field potential~(\ref{pot}) were chosen as
$n=2$, $\lambda =1$, and $M=10^{-2}m_p$.
The resulting solutions for $f(r)$, $A(r)$, and $B(r)$ are shown in 
Figs.~\ref{fig=gbmf} and~\ref{fig=gbm}. 
The solution for the scalar field $f(r)$ is not
much different from the flat space solution with the same values 
of the parameters.
Outside the monopole core, the metric coefficient $A(r)$ is very accurately
given by
\begin{equation}
A^{-1}(r)=1-\kappa\lambda^{1/2} GM^3r\;,
\label{mA}
\end{equation}
where $\kappa = 25.6$. 
Note that it is in a good agreement with the linearized theory
which gives $\kappa = kK =17\pi/2$.
From Eq.~(\ref{mA}) we see that
$A(r)$ diverges at $r=r_0= (\kappa\lambda^{1/2} GM^3)^{-1}$.
$B(r)$ vanishes at the same point, while the product $AB$ remains finite.
The singularity at $r=r_0$ is only a coordinate singularity.
It is of the same type as the singularity at the horizon,
$r=H^{-1}$, in the static de Sitter metric,
\begin{equation}
ds^2=-(1-H^2r^2)dt^2+(1-H^2r^2)^{-1}dr^2 + r^2d\Omega^2\;.
\end{equation}
An observer at the monopole center will see test particles accelerating
away and ``fading away'' as they approach the horizon.

We note that $f(r_0)/m_p=0.18\sim 1$. This indicates that the horizons form
when the field $f$ becomes comparable to $m_p$ in the space between
monopoles, that is, when the typical separation between monopoles
is $R\sim r_0$. It was argued in Ref.~\cite{Cho} that the monopoles
start dominating the universe at about the same time.
We have verified that the value of $f(r_0)/m_p$ is roughly independent
of $M$.

For a gauge monopole, the gravitational field at $r\ll R$ is determined
mainly by the energy concentration near the core, while
the potential energy term is negligible.
In flat spacetime we have found~\cite{Cho} that the potential $V(f)$
has a negligible effect on the scalar field as well,
which corresponds to the Prasad-Sommerfield (PS) limit.
The role of the potential is only to determine the constant $a$
appearing in the PS solution~(\ref{fggm}).
We expect the situation in curved spacetime to be similar.
As we discussed in~\cite{Cho}, when $R$ is increased, $a$ grows,
the monopole mass grows, and the size of its core decreases,
so that the monopole core becomes a black hole at some critical
value $a=a_c$.
Gravitating monopoles in the PS limit have been studied numerically
by Ortiz~\cite{Ortiz} and by Breitenlohner, Forg\'acs and 
Maison~\cite{Maison}.
They found $a_c \approx 0.4m_p$. We obtained numerical solutions
for vacuumless gauge monopoles and found essentially identical results.
We shall not reproduce them here.

For a global string, we used the ansatz~(\ref{smetric}) with the
boundary conditions
\begin{eqnarray}
B(0)=1\;, \quad C(0)=0\;,\nonumber\\
B'(0)=C'(0)=0\;,
\label{BCBC}
\end{eqnarray}
and Eq.~(\ref{BCf}).

\begin{figure}
\psfig{file=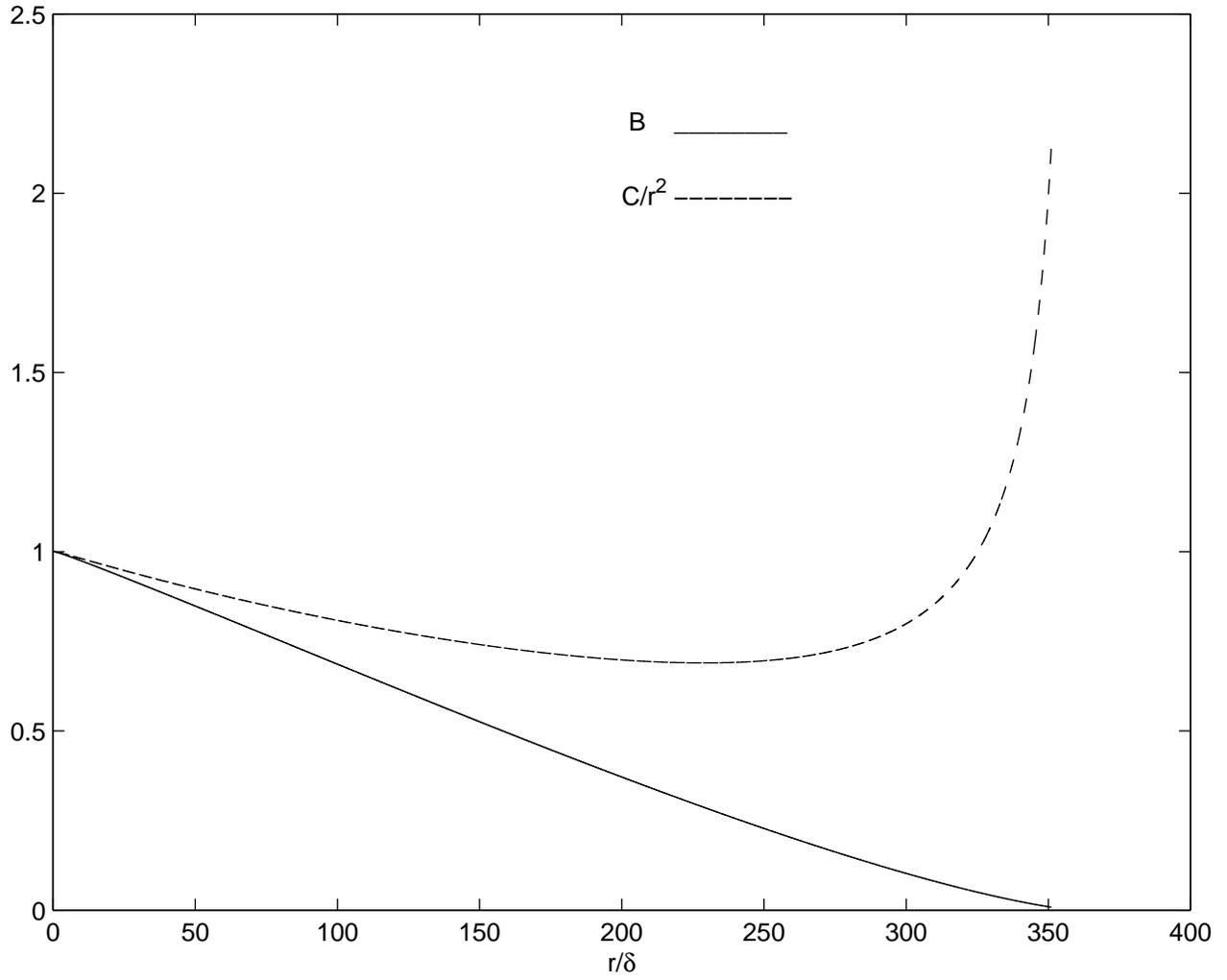}
\caption{
Metric coefficients $B(r)$ and $C(r)/r^2$ of a vacuumless global string
with $M=0.01m_p$.
}
\label{fig=gbs}
\end{figure}

\begin{figure}
\psfig{file=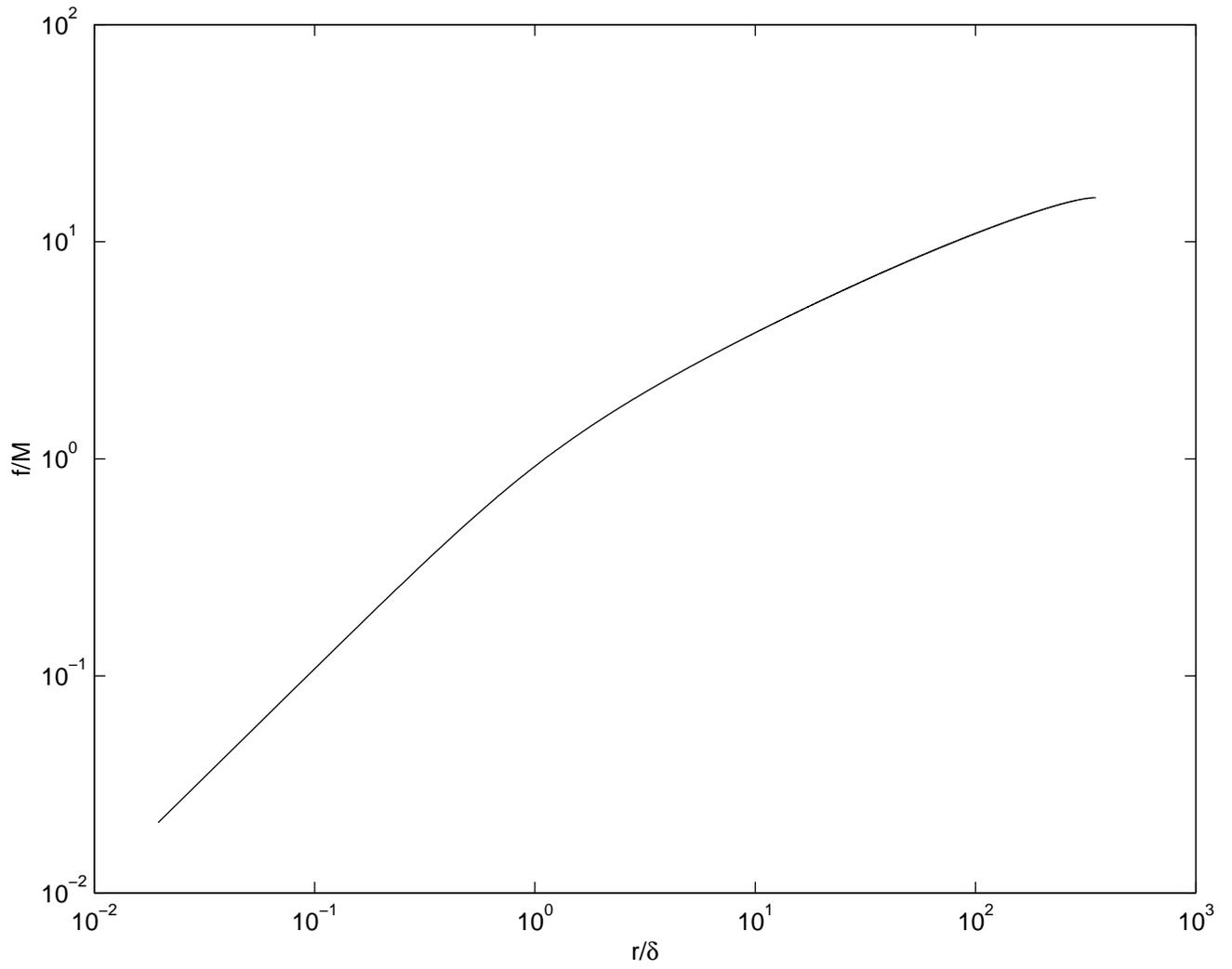}
\caption{
Scalar field $f(r)$ of a vacuumless global string
with $M=0.01m_p$.
}
\label{fig=gbsf}
\end{figure}

The numerically calculated functions $B(r)$ and $C(r)/r^2$ are
plotted in Fig.~\ref{fig=gbs}. As for a global monopole, $B(r)$ vanishes
at a finite distance from the core, $r=r_0$, and
$C(r)$ diverges on the same surface.
However, in the case of string this appears to be a true spacetime
singularity, since the invariant $R_{\mu\nu\sigma\tau}R^{\mu\nu\sigma\tau}$
diverges at $r\to r_0$.
The scalar field $f(r)$, shown in Fig.~\ref{fig=gbsf}, appears to be
well-behaved at $r\to r_0$.
We find that $f(r_0)=0.16m_p$ and is essentially independent of $M$
(for $M \ll m_p$). 
Hence, singularities do not form until $f$ becomes comparable to $m_p$
in the space between the strings.

The singularity of a global vacuumless string spacetime is not
so surprising if we recall that static spacetimes of ``ordinary''
global strings exhibit similar singularities~\cite{Cohen}.
Gregory~\cite{Gregory} has recently argued that the singularity can be 
removed if the space is allowed to expand along the string axis
in a de Sitter-like fashion.
The same argument may apply to vacuumless strings.

\begin{figure}
\psfig{file=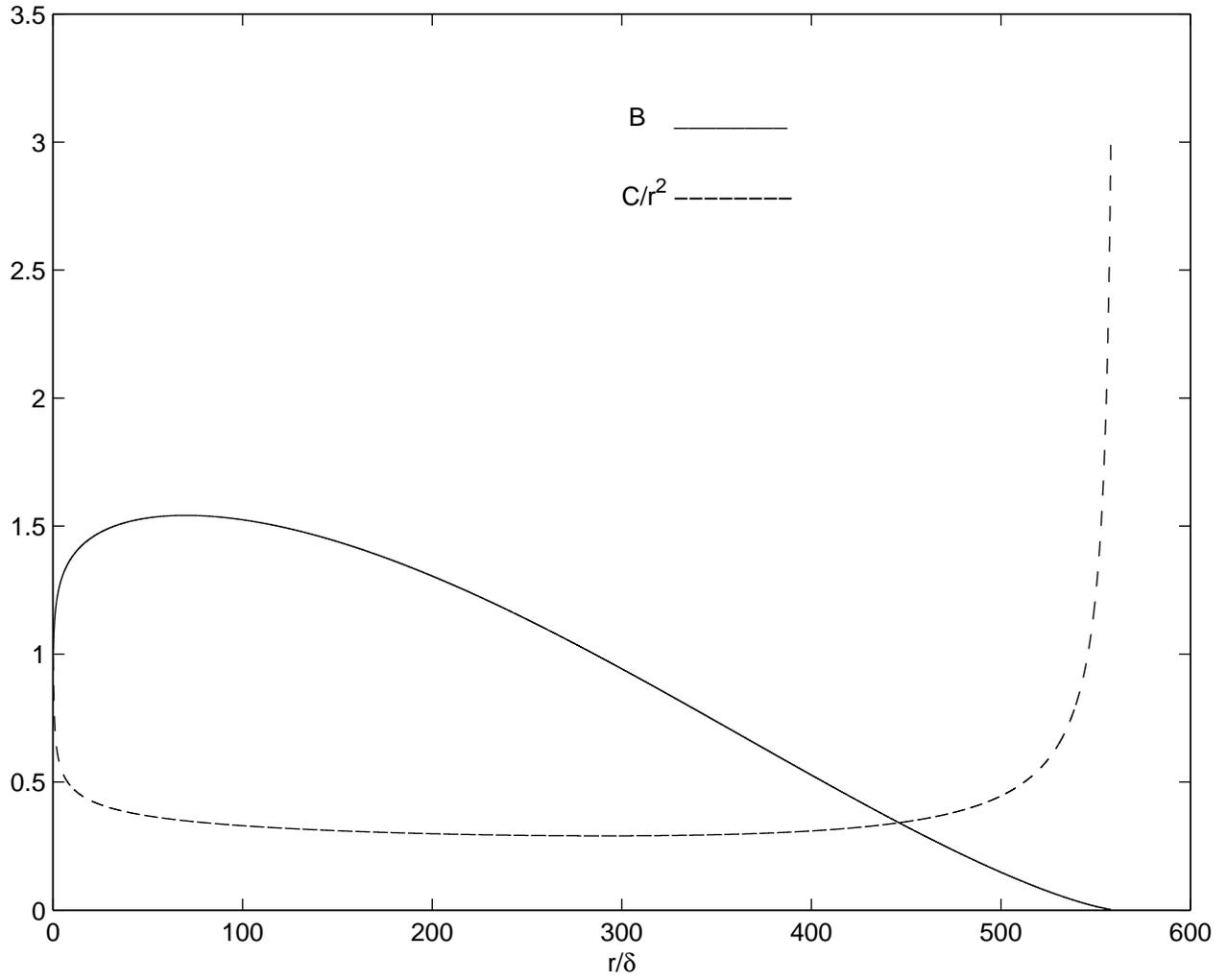}
\caption{
Metric coefficients $B(r)$ and $C(r)/r^2$ of a vacuumless gauge string
with $M=0.02m_p$ and $\lambda /e^2=1$.
}
\label{fig=ggs}
\end{figure}

We finally consider a vacuumless gauge string. The functions $B(r)$
and $C(r)/r^2$ for such a string are shown in Fig.~\ref{fig=ggs}.
At small values of $r$ they agree well with the linearized
solution~(\ref{lrgsmetric}) describing an attractive gravitational
potential. The character of the metric changes when the potential
$V(f)$ becomes comparable to the gradient terms in $T_{\mu\nu}$.
At large $r$ the potential dominates and the gravitational field of
the string is repulsive.

As the cutoff radius $R$ is increased, $f(R)$ grows.
When $f(R)$ approaches the critical value $f_c\approx 0.66m_p$ (roughly
independent of $M$ for $M\ll m_p$), $B(r)$ tends to zero, while
$C(r)$ begins to diverge. The curvature invariant
$R_{\mu\nu\sigma\tau}R^{\mu\nu\sigma\tau}$ also diverges, 
indicating a true spacetime singularity.
The solutions for $f(r)$ and $\alpha (r)$ are shown in
Fig.~\ref{fig=ggsf}. (The gauge field $\alpha (r)$ is defined in the
Appendix.)

To elucidate the nature of this singularity, we attempted to
approximate the gauge string metric at large $r$ analytically.
The flat-space solution~(\ref{fggs}) suggests that $f(r)$ is
a slowly-varying function, so one can expect that the potential
$V(f)$ will be nearly constant for $r\sim R$.
Indeed, numerically we find that $V$ changes very little in this
range. An approximate representation of the string metric can then be
found by solving Einstein's equation with the cylindrically-symmetric
ansatz~(\ref{smetric}) and with a constant vacuum energy source,
$V=V_0$. The solution is
\begin{equation}
ds^2=\cos^{4/3}(Qr)(-dt^2+dz^2)+dr^2+Q^{-2}\tan^2(Qr)\cos^{4/3}(Qr)
d\theta^2\;,
\end{equation}
where $Q=(32\pi V_0/3m_p^2)^{1/2}$.
It has a geometric singularity at $r=\pi/2Q$.  The qualitative
behavior of this metric near the singularity is similar to
that of our numerical solution.

\section{Conclusions}
We have studied the gravitational fields produced by vacuumless
strings and monopoles. For a global monopole, the field is strongly
repulsive and the spacetime has an event horizon similar to that in
de Sitter space. Only the interior of the horizon is describable
by a static metric.

The metric of a gauge vacuumless monopole is very similar to that of
ordinary monopole in the Prasad-Sommerfield limit.
When the Higgs field outside the monopole reaches a critical value
$f_c\sim m_p$, the spacetime develops an event horizon and the monopole
core becomes a black hole.

For a global string, the gravitational field is strongly repulsive
and the spacetime becomes singular at a finite distance from the
string core.

The gravitational field of a gauge string is attractive at small
distances and repulsive at large distances. There is also a spacetime
singularity at a finite distance from the core.

The singular nature of the string spacetimes is a cause for some concern.
As we argued in~\cite{Cho}, vacuumless strings can naturally arise
at a phase transition in the early universe. 
If the evolution of such strings leads to naked singularities,
we would have a contradiction with the cosmic censorship 
hypothesis~\cite{Penrose} which forbids the formation of naked singularities
from regular initial conditions.
Once naked singularities are formed, the subsequent evolution of
the universe can no longer be determined.
No region of space would be immune to this problem, since the string
network permeates the entire universe.
Hence, the distance from a typical observer to a singularity
would be comparable to the average distance between the strings.

A possible way out of this difficulty may be to lift the requirement
that the vacuumless string metric should be static.
Static spacetimes of ordinary global strings also exhibit 
singularities~\cite{Cohen}, and Gregory~\cite{Gregory} has argued that
these singularities can be removed if the space is allowed to 
expand in a de Sitter-like fashion along the string axis.
(The situation would then be similar to the nonsingular spacetime of
a domain wall~\cite{VilenkinW,Ipser} 
which is expanding in the plane of the wall.)
This issue requires further study.

\acknowledgements
This work was supported in part by the National Science Foundation.

\appendix
\section*{Field equations}
For strings, with the Lagrangian~(\ref{lag}) and the 
metric~(\ref{smetric}) Einstein's equations are
\begin{equation}
{1 \over 2}{B'' \over B}+{1 \over 4}{B' \over B}{C' \over C}
=8\pi G(T_t^t-{1 \over 2}T)\;,
\end{equation}
\begin{equation}
{B'' \over B}+{1 \over 2}{C'' \over C}
-{1 \over 2}\left({B' \over B}\right)^2
-{1 \over 4}\left({C' \over C}\right)^2
=8\pi G(T_r^r-{1 \over 2}T)\;,
\label{Grr}
\end{equation}
\begin{equation}
{1 \over 2}{C'' \over C}+{1 \over 2}{B' \over B}{C' \over C}
-{1 \over 4}\left({C' \over C}\right)^2
=8\pi G(T_\theta^\theta-{1 \over 2}T)\;,
\end{equation}
where
\begin{equation}
T_t^t=T_z^z={f'^2 \over 2}
+{f^2(1-\alpha )^2 \over 2C}+{\alpha '^2 \over 2e^2C}+V(f)\;,
\end{equation}
\begin{equation}
T_r^r=-{f'^2 \over 2}
+{f^2(1-\alpha )^2 \over 2C}-{\alpha '^2 \over 2e^2C}+V(f)\;,
\end{equation}
\begin{equation}
T_\theta^\theta={f'^2 \over 2}
-{f^2(1-\alpha )^2 \over 2C}-{\alpha '^2 \over 2e^2C}+V(f)\;,
\end{equation}
and $T=T_\mu^\mu$.
With the string ansatz for the gauge field, 
$A_\theta (r)=-\alpha (r)/er$, 
the scalar and gauge field equations are 
\begin{equation}
f''+\left( {B' \over B} + {C' \over 2C}\right)f'
-{1 \over C}f(1-\alpha)^2-{\partial V(f) \over \partial f}=0\;,
\end{equation}
\begin{equation}
\alpha '' +\left( {B' \over B} - {C' \over 2C}\right)
\alpha ' +e^2f^2(1-\alpha)=0\;.
\end{equation}
The equations with $\alpha =0$ are for the global string.
We numerically solve the above equations with the boundary 
conditions~(\ref{BCf}), (\ref{BCBC}),
$\alpha (0)=0$, and $\alpha (R)=1$.

For monopoles, with the metric~(\ref{mmetric}) and the gauge field
$A^a_i = -[1-\omega (r)]\epsilon^{aij}x^j/er^2$
Einstein's equations are
\begin{equation}
{A' \over A^2r}-{1 \over Ar^2}+{1 \over r^2}
=8\pi G \left[ {f'^2 \over 2A} + {f^2\omega^2 \over r^2}
+{1 \over e^2r^2} \left[ {\omega '^2 \over A}
+{(1-\omega^2)^2 \over 2r^2} \right] +V(f) \right]\;,
\end{equation}
\begin{equation}
{(AB)' \over AB} = 16\pi Gr \left( {f'^2 \over 2}
+{\omega '^2 \over e^2r^2} \right)\;,
\end{equation}
and the field equations are
\begin{equation}
{1 \over A}f''+{1 \over A}\left( {B' \over 2B} - {A' \over 2A}
+{2 \over r^2} \right)f'
-{2 \over r^2}\omega^2f
-{\partial V(f) \over \partial f}=0\;,
\end{equation}
\begin{equation}
{1 \over A}\omega '' +{1 \over 2A} \left( {B' \over B} - {A' \over A}\right)
\omega ' + {\omega (1-\omega^2) \over r^2} -e^2f^2\omega =0\;.
\end{equation}
The equations with $\omega =0$ are for the global monopole.
We apply boundary conditions~(\ref{BCm}), (\ref{BCf}),
$\omega (0)=1$, and $\omega (R) =0$.

\begin{figure}
\psfig{file=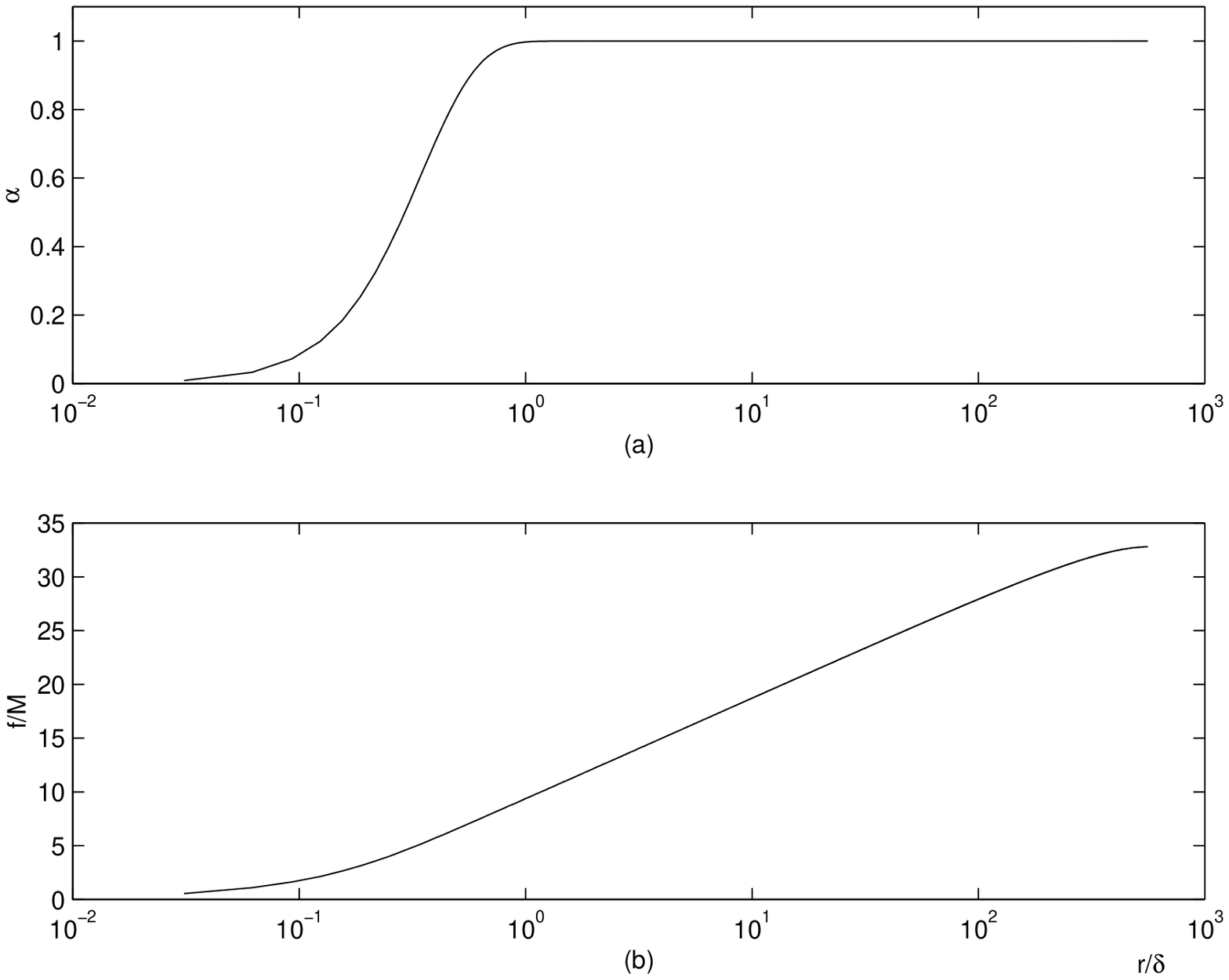}
\caption{
(a) Gauge field $\alpha (r)$ and (b) scalar field $f(r)$ 
of a gauge vacuumless string
with $M=0.02m_p$ and $\lambda /e^2=1$.
}
\label{fig=ggsf}
\end{figure}

\end{document}